\documentclass[graybox]{svmult}

\usepackage{mathptmx}       % selects Times Roman as basic font
\usepackage{helvet}         % selects Helvetica as sans-serif font
\usepackage{courier}        % selects Courier as typewriter font
\usepackage{type1cm}        % activate if the above 3 fonts are
\usepackage{makeidx}         % allows index generation
\usepackage{graphicx}        % standard LaTeX graphics tool
\usepackage{multicol}        % used for the two-column index
\usepackage[bottom]{footmisc}% places footnotes at page bottom

\usepackage{amsmath} 
\usepackage{amssymb} 
\newcommand{\bmat}{\left[\begin{matrix}}
\newcommand{\emat}{\end{matrix}\right]}
\makeindex             % used for the subject index
\begin{document}

\title*{Prediction of Pedestrian Speed with Artificial Neural Networks}
\author{Antoine Tordeux, Mohcine Chraibi, Armin Seyfried and Andreas Schadschneider}
\institute{Antoine Tordeux \at Forschungszentrum J\"ulich and University of Wuppertal, \email{a.tordeux@fz-juelich.de}
\and Mohcine Chraibi \at Forschungszentrum J\"ulich, \email{m.chraibi@fz-juelich.de}
\and Armin Seyfried \at Forschungszentrum J\"ulich and University of Wuppertal, \email{a.seyfried@fz-juelich.de}
\and Andreas Schadschneider \at University of Cologne, \email{as@thp.uni-koeln.de}}
\maketitle

\abstract{Pedestrian behaviours tend to depend on the type of facility.
  Therefore accurate predictions of pedestrians movements in complex
  geometries (including corridor, bottleneck or intersection) are
  difficult to achieve for classical models with few parameters. 
	Artificial neural networks have multiple parameters and are able to 
	identify various types of patterns. They could be a 
	suitable alternative for forecasts. We aim in this paper to present first steps 
	testing this approach.  We compare estimations of pedestrian speed 
	with a classical model and a neural network for combinations of 
	corridor and bottleneck experiments.  
	The results show that the neural network is able to
  differentiate the two geometries and to improve the estimation of
  pedestrian speeds when the geometries are mixed.}

\section{Introduction}
\label{sec:1}
%\begin{equation}gin{svgraybox}Existence of phase transitions for the stop-and-go dynamics in real traffic flow remains controversially\end{svgraybox}

Microscopic pedestrian models are frequently used in traffic
engineering to predict crowd dynamics.  Classical operational
approaches are in general decision-based, velocity-based or force-based models (see
\cite{Schadschneider2009} and references therein).  Such models consider 
physical as well as social or psychological factors.
They are basic rules or generic functions depending locally on the
environment. Few parameters having generally physical interpretations 
allow to adjust the model. 

Before applying simulations to make predictions, the model parameters 
have to be calibrated and the models have to be validated, experimentally or statistically by using real data. 
The validation can be carried out by checking whether the models are able to describe 
the dynamics accurately for configurations different from the ones used for the calibration 
(cross-validation) \cite{Treiber2013}. 
A good model should provide realistic dynamics in different conditions 
(i.e.\ different geometries, initial or boundary conditions) for the same 
setting of the parameters.

The parameter for the convection part of the models (for instance desired speed or time gap) 
can be referred to the fundamental diagram (FD), a phenomenological
relation between speed and surrounding distance spacing to the neighbours and obstacles \cite{Seyfried2005}. 
This relation can be explicitly used to model the speed of the pedestrian 
and is then related to \emph{optimal velocity}, a concept 
borrowed from traffic modelling \cite{Bando1995}, see
e.g.\ \cite{Nakayama2005,Moussa2012,Lv2013}. 
It is also used as an implicit
relation (see e.g.\ \cite{Helbing1995,Chraibi2010,GUO2010}) that is 
determined by considering uni-dimensional flows \cite{Chraibi2015}. 
By neglecting anisotropic effects in the models (such as the vision based effect), 
microscopic models can be characterised at an aggregated level by fundamental diagrams determining a speed 
to a local density given by the mean distance spacing to the closest neighbours \cite{Das2015}. 
In the following we refer a model simply based on a fundamental diagram as {\em FD-based model}.
%One refers in the following as \emph{FD-based model} a model simply based on a fundamental diagram relation. 

Despite of their simplicity, microscopic models can describe realistic
pedestrian flows, as well as self-organization phenomena such as lane formation or 
alternation of flow at a bottleneck in bi-directional streams \cite{helbing2005,Schadschneider2009}.  
However, the prediction of
pedestrian movement in complex spatial structures (e.g.\ 
buildings like stadia or stations) remains problematic. Observations show that pedestrians
tend to adapt their behaviour according to the facilities
\cite{Daamen2004}.  For instance, the flow tends to locally increase
at bottlenecks \cite{Seyfried2009,Jun2014,PIP347}.  This
leads to geometry-dependent characteristics and makes aggregated models
based on a single fundamental diagram unable to accurately describe transitions
between different types of facilities (such as corridor, T-junction,
crossing or bottleneck), as well as from platforms to stairs.

\begin{figure}[!ht]
\includegraphics[width=1\textwidth]{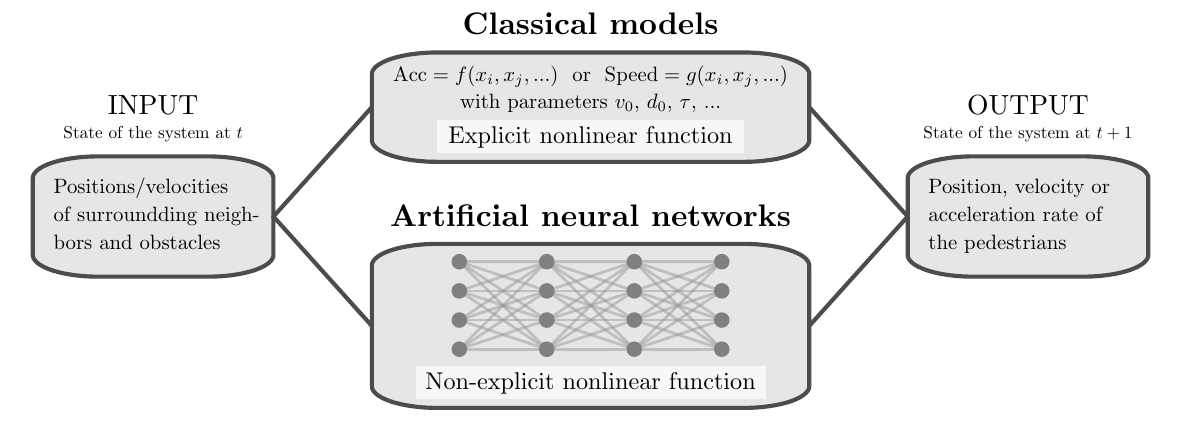}
\caption{Minimalistic illustrative scheme for the distinction between FD-based models, which are explicit non-linear 
functions calibrated by few meaningful parameters, 
and neural networks, for which the non-linear function is data-based and has deliberately a large number of parameters.}
\label{fig:1} \vspace{-4mm}      % Give a unique label
\end{figure}

An alternative data-driven approach for the prediction of pedestrian
dynamics in complex geometries could be provided by using artificial neural
networks (ANN).  Neural networks have already proven their efficiency for motion
planning in robotic or autonomous vehicles
\cite{sadati2002,Jackel2007}, and start to be used for pedestrian
dynamics as well \cite{Das2015,Fragkiadaki2015,Ma2016,Alahi2016}.  Such approach is
data-based and, by opposition to classical models, has artificially a
very large number of parameters with no explicit physical meaning (see
Fig.~\ref{fig:1}).  ANN can reproduce complex patterns
that FD-based models, describing dynamics at an aggregated level,
cannot.

The aim of this work is to evaluate whether ANN could accurately 
describe pedestrian behaviour for different types of geometries. 
A feed-forward neural network is compared to a FD-based model
with data gained by experiments at bottleneck and corridor with closed boundary conditions 
(in the following `bottleneck' and `ring' experiments) \cite{zenodo,data}. 
The performances (i.e.\ the fundamental diagram) significantly differ
according to the spatial structure.  Training and testing (cross-validation)
are carried out for different combinations of bottleneck and ring
experiments.  The results show that the neural network is able to
identify the spatial structure of the facility and improve the prediction in case of mixed
structures. 
%The paper is organised as following. 
The data and the models used are
presented in Secs.~\ref{sec:2} and \ref{sec:3}.  The fitting of the
neural network is proposed in Sec.~\ref{sec:4} while the comparison to
the FD-based model over combinations of bottleneck and ring
experiments are given in Sec.~\ref{sec:5}.  Conclusions are
presented in Sec.~\ref{sec:6}.

\section{Models}
\label{sec:2}
The pedestrian modelling approaches are continuous speed models based
on the $K=10$ closest neighbours.  We denote in the following $(x,y)$
as the position of the considered agent, $v$ as its speed and
$\big((x_i,y_i),i=1,\ldots,K\big)$ as the positions of the $K$ closest
neighbours.

The first modelling approach is the Weidmann model for the fundamental
diagram \cite{Weidmann1993} for which the speed is a function of the
mean spacing (i.e.\ the local density) %Model for understanding:
\begin{equation}
v=\text{FD}(\bar s_K,v_0,T,\ell)=v_0\Big(1-e^{\frac{\ell-\bar s_K}{v_0T}}\Big).
\end{equation}
Here $\bar s_K=\frac1K\sum_{i=1}^K\sqrt{(x-x_i)^2+(y-y_i)^2}$ is the mean
spacing with the $K=10$ closest neighbours while the time gap $T$, the
pedestrian size $\ell$ and the desired speed $v_0$ are the parameters
of the model.

The second modelling approach is a feed-forward artificial neural
network with hidden layers $H$ %Models for prediction:
%\vspace{4mm}\be\left|\begin{array}{l}~~v=\text{NN}_1\big(H_1,\sqrt{(x_i-x)^2+(y_i-y)^2},1\le i\le K\big)\\[4mm]~~v=\text{NN}_2\big(H_2,x_i-x,y_i-y,1\le i\le K\big)\end{array}\right.
\begin{equation} v=\text{NN}\big(H,\bar s_K,(x_i-x,y_i-y,1\le i\le K)\big).
\end{equation}
The network has $2K+1$ inputs: The mean spacing and the $K$ relative
positions.  The number of parameters in the network depends on the
number of nodes in the hidden layers.

\section{Data}
\label{sec:3}
Two datasets obtained in laboratory conditions are used to compare the
FD-based and ANN modelling approaches. 
The data are available on the internet (see \cite{zenodo}). 
They are part of the online database of pedestrian experiments \cite{data}. 
The first dataset,
denoted R and called the ring experiment, comes from a experiment done on a closed geometry of length 30~m and width 1.8~m for
different density levels (ranging from 0.25 to 2~ped/m$^2$ --- the participant number ranges from 15 to 230).  The
second dataset, denoted B, is an experiment for a bottleneck geometry.
The width of the system in front of the bottleneck is 1.8~m while the width
of the bottleneck varies (from 0.70, 0.95, 1.20 to 1.80~m --- 150 participants by experiment). 
See \cite{zenodo} for details on the data. The
flow and density are measured every 10~s to deal with pseudo-independent
measurements.  Each sample contains around $N=2,000$ observations.

\begin{figure}[!ht]
\sidecaption
\includegraphics[width=7.5cm]{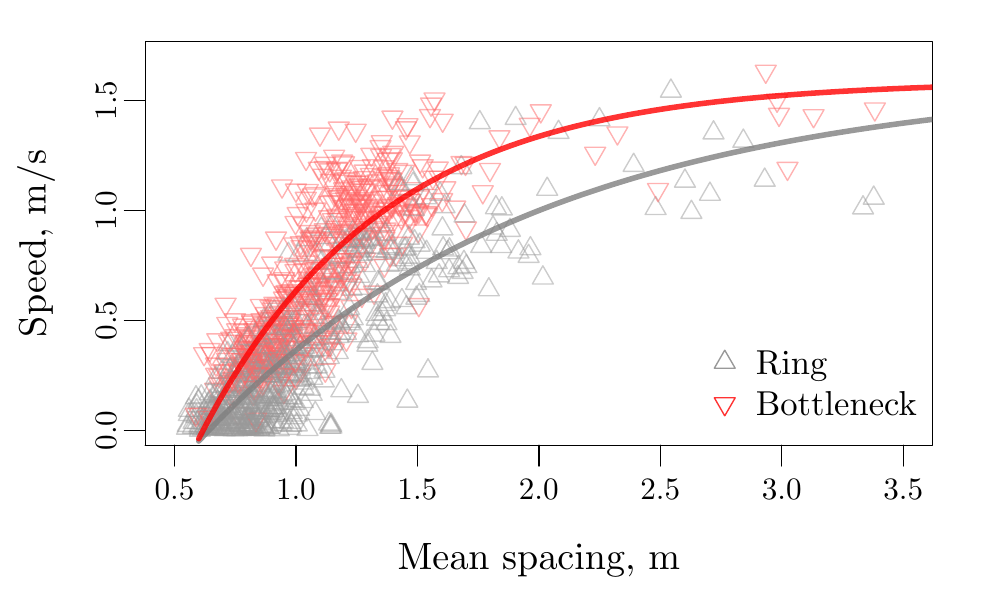}
\caption{Observations of the pedestrian speeds as function of the mean spacing with the 10 closest neighbours 
for the ring and bottleneck experiments. Two distinct relationships can be identified.}
\label{fig:2}       % Give a unique label
\end{figure}

The two data sets describe two different interaction behaviours (see
Fig.~\ref{fig:2}).  The speed for a given mean spacing tends to be
higher in the bottleneck than on the ring when the system is
congested.  Consequently the
estimation of the time gap significantly differs according to the
geometry (see Table~\ref{tab:1}).

\begin{table}
%\sidecaption
\caption{Fitting of the time gap $T$, the pedestrian size $\ell$ and 
the desired speed $v_0$ parameters of the Weidmann model on the ring and bottleneck experiment. 
The time gap significantly differs according to the geometry. }
\begin{tabular}{p{2cm}p{2cm}p{2cm}}
\hline\noalign{\smallskip}
Experiment&R&B\\
\noalign{\smallskip}\svhline\noalign{\smallskip}
$\ell$ (m) & 0.64&0.61\\
$T$ (s)& 0.86&0.48\\
$V_0$ (m/s)& 1.60&1.58\\
\noalign{\smallskip}\hline\noalign{\smallskip}
\end{tabular}
\label{tab:1}  
\end{table}

\section{Fitting the neural network}
\label{sec:4}

The neural network is fitted with cross-validation and bootstrap \cite{Mooney1993,Kohavi95} over
50 sub-samples.  The training is performed using half of the data while the
network is tested on the remaining.  The training is carried out with the
back-propagation method \cite{Rumelhart1986} on the normalised data, by minimising from the
top to the bottom of the network the mean square error
\begin{equation}
\text{MSE}=\frac1N\sum_i\big(v_i-\tilde v_i\big)^2,
\end{equation}
with $v_i$ the observed speed, $\tilde v_i$ the predicted speed and
$N$ the number of observations.  The computation is done with the
statistical software R \cite{RR} and the package \texttt{neuralnet} \cite{neuralenet}.

\begin{figure}[!ht]
\sidecaption
\includegraphics[width=7.5cm]{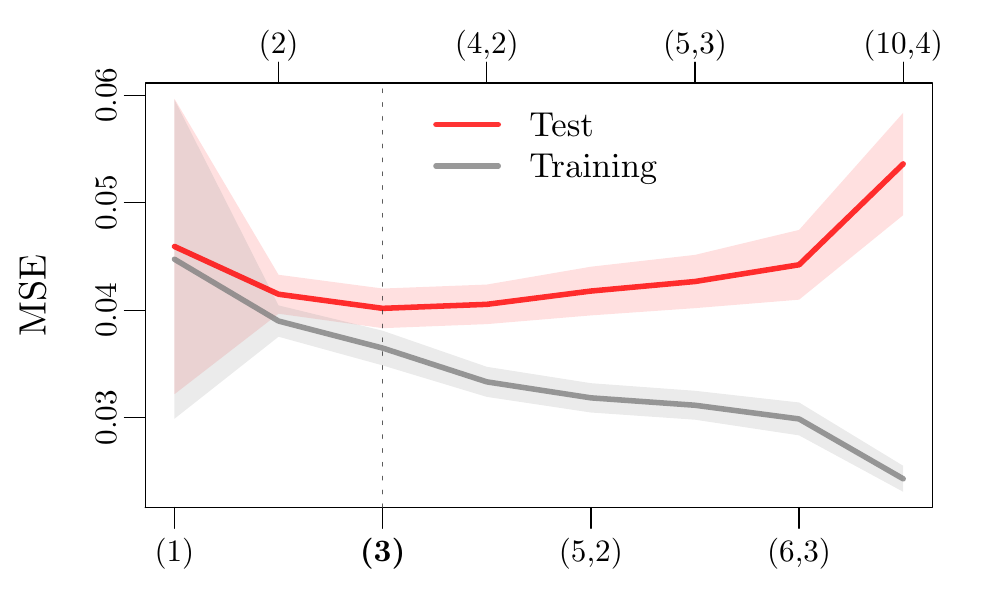}
\caption{Training and testing errors according to different hidden in
  the network.  The curves correspond to the average of 50-bootstrap
  estimates while the bands describe the standard deviation.  The
  training error systematically decreases as network complexity
  increases while the testing error admits a minimum for hidden $H=$
  (3).}
\label{fig:3}       % Give a unique label
\end{figure}

The different hidden layers (1), (2), (3), (4,2), (5,2), (5,3), (6,3),
(10,4) and (12,5) are tested (see Fig.\ref{fig:3}).  As expected, the
training error tends to decrease as the complexity of the network
increases, while the testing error shows a minimum before overfitting. 
Such a minimum is reached for the
single hidden layer $H=$ (3) with 3 nodes.  Note that here the
calibration is done on a combination of the ring and bottleneck
experiment datasets.  Yet similar results are obtained when
calibrating separately on the ring and on the bottleneck datasets.

\section{Model comparison}
\label{sec:5}

\begin{figure}[!ht]
\sidecaption
\includegraphics[width=7.5cm]{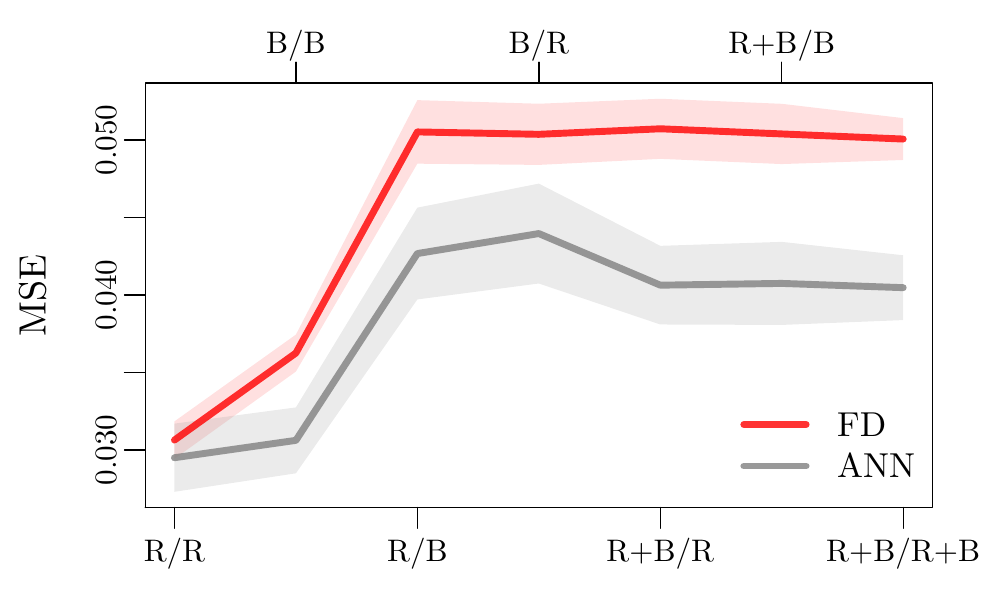}
\caption{Testing error of the FD-model and the neural network according to combinations of the ring R and bottleneck B experiments. 
The curves correspond to the average of 50-bootstrap estimates while the bands describe the standard deviation. 
The improvement of the speed is significant by using the network when the experiments are mixed (i.e.\ R+B). }
\label{fig:4}  % Give a unique label
\end{figure}

\begin{figure}[!ht]
\sidecaption
\includegraphics[width=7.5cm]{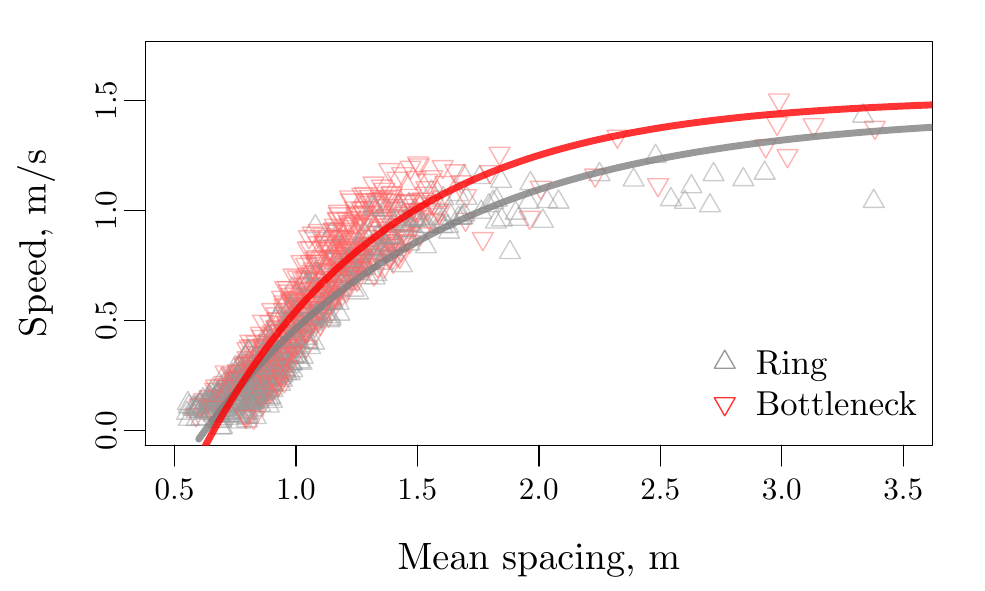}
\caption{Prediction by the neural network of the pedestrian speed for the R+B$/$R+B training and testing datasets. 
The network is able to, at least partially, identify the two geometries. As observed in the real data, the speed for a given mean spacing is in average in the bottleneck higher than the flow in the corridor for congested situations.}
\label{fig:5}      % Give a unique label
\end{figure}

The calibrated FD-model and the trained neural network with $H=3$ are
compared on different combinations of data of the ring R and bottleneck B
experiments. Here the first argument in the notation `.\,/\,.'\ corresponds to the
training phase, while the second argument corresponds to the testing phase. 
The testing errors are presented in Fig.~\ref{fig:4}. 
The modelling approaches are firstly analysed on the identical sets
R$/$R and B$/$B.  Here the network is slightly better than the
FD-model.  For the ring experiment, the performances are relatively
homogeneous and the MSE is only approximately 5\% smaller by using the
network.  While for the bottleneck, the performances fluctuate more 
and the improvement is more significant (around 15\%).  The
improvement is also significant when the approaches deal with
unobserved situations, i.e. for the datasets R$/$B and B$/$R (around
15\%).  The best results are obtained when training the models on the
combination of ring and bottleneck experiments, i.e.\ the scenarios
R$/$R+B, B$/$R+B and R+B$/$R+B.  As shown in Fig.~\ref{fig:5} and
Table~\ref{tab:2}, the network is able in such situation to partially
differentiate the two geometries and to improve the speed estimation
by a factor of approximately 20\%.  The orders of improvement are
similar to the ones obtained in \cite{Alahi2016} with the social LSTM
neural network and the social force pedestrian model \cite{Helbing1995}.
%Such proportion is roughly the difference we observe in the speed for the two experiments. 

\begin{table}
\caption{Fitting of the time gap $T$, the pedestrian size $\ell$ and 
the desired speed $v_0$ parameters for the data predicted by the neural network.}
\begin{tabular}{p{2cm}p{2cm}p{2cm}}
\hline\noalign{\smallskip}
Experiment&R&B\\
\noalign{\smallskip}\svhline\noalign{\smallskip}
$\ell$ (m) & 0.63&0.66\\
$T$ (s)& 0.68&0.50\\
$V_0$ (m/s)& 1.44&1.51\\
\noalign{\smallskip}\hline\noalign{\smallskip}
\end{tabular}
\label{tab:2}
\end{table}

\section{Conclusion}
\label{sec:6}

The data-driven approach using an artificial neural network is able to distinguish pedestrian performances in ring and bottleneck experiments 
from the relative positions of the $K=10$ closest neighbours and the mean spacing. 
Consequently, we observe that the speed prediction for mixed data can be improved by a factor up to 20\% by using a network 
compared to an aggregated model based on fundamental diagrams.

The results are first steps suggesting that neural networks could be
suitable tools for the prediction of pedestrian dynamics in complex
geometries.  Yet, the simulation of the networks remain to be carried out
over full trajectories and compared to the performances obtained with
existing models and notably anisotropic models.  Furthermore, other inputs, hidden layers and
training on different geometries have to be investigated.  Especially,
one remains to test the complexity necessary to the network for
accurate precisions regarding to the size and heterogeneity of the
datasets.
%Note that the computation times are relatively important (several hours on a 2.7~Ghz processor for the 50-bootstrap training of the (5,3) layer neural network over $n=$ 2,000 observations). Therefore for large database and complex networks the use of super computers seems necessary. 

\begin{acknowledgement}
  %The authors would like to thank Alessandro Corbetta for valuable discussions on the setting of neural networks. 
	Financial supports
by the German Science Foundation (DFG) under grants SCHA 636/9-1 and SE 1789/4-1 are gratefully
acknowledged.
\end{acknowledgement}

%The reference style for working with bibtex is 

\end{document}